\documentclass[aps,prd,superscriptaddress,10pt,nofootinbib,preprintnumbers]{revtex4} 

\usepackage[utf8]{inputenc} 

\usepackage{graphicx}
\usepackage{dcolumn}
\usepackage{bm}
\usepackage{latexsym}
\usepackage{amsfonts}
\usepackage{amssymb}
\usepackage{amsmath}
\usepackage{color}

\begin{document}

\preprint{Imperial/TP/2016/AEG/3, YITP-16-66, IPMU16-0079} 

\title{Jeans' Ghost}

\author{A. Emir G\"umr\"uk\c{c}\"uo\u{g}lu}
\affiliation{Theoretical Physics Group, Blackett Laboratory,
Imperial College London, South Kensington Campus, London, SW7 2AZ,
UK}

\author{Shinji Mukohyama}
\affiliation{Center for Gravitational Physics, Yukawa Institute for Theoretical Physics, Kyoto University, 606-8502, Kyoto, Japan}
\affiliation{Kavli Institute for the Physics and Mathematics of the Universe (WPI), The University of Tokyo Institutes for Advanced Study, The University of Tokyo, Kashiwa, Chiba 277-8583, Japan}

\author{Thomas P. Sotiriou}
\affiliation{School of Mathematical Sciences, University of Nottingham, University Park, Nottingham, NG7 2RD, UK}
\affiliation{School of Physics and Astronomy, University of Nottingham, University Park, Nottingham, NG7 2RD, UK} 

\date{\today}
\begin{abstract}
We show that  a massless canonical scalar field  minimally coupled to general relativity can become a tachyonic ghost at low energies around a background in which the scalar's gradient is spacelike. By performing a canonical transformation we demonstrate that this low energy ghost can be recast, at the level of the action, in a form of a fluid that undergoes a Jeans--like instability affecting only modes with large wavelength. This illustrates that low-energy tachyonic ghosts do not lead to a catastrophic quantum vacuum instability, unlike the usual high-energy ghost degrees of freedom. 
\end{abstract}
\maketitle

\section{Introduction}
\label{sec:introduction}

Motivated by the observation of accelerated expansion \cite{Riess:1998cb,Perlmutter:1998np}, there has been an increasing interest towards alternative gravity theories with the goal of sourcing the late time acceleration by introducing an IR modification to general relativity (GR).\footnote{See Ref.~\cite{Clifton:2011jh} for a review.} However, apart from a handful of exceptions (e.g. unimodular gravity \cite{Unruh:1988in,Henneaux:1989zc,Kluson:2014esa}), these theories contain additional dynamical degrees of freedom on top of the two tensor modes of GR. The stability of the extra modes around flat, cosmological and spherically symmetric backgrounds is a powerful way in determining the consistency of the theories. Particularly, one often comes across ghost modes around these backgrounds, e.g. the self-accelerating branch of DGP \cite{Luty:2003vm}, linear dS massive gravity \cite{Higuchi:1986py}, generic nonlinear massive gravity \cite{Boulware:1973my} to name a few. 

In the context of field theory, a ghost is a degree of freedom which has a negative kinetic energy (see Ref.~\cite{Sbisa:2014pzo} for a review). Consider the Lagrangian
\begin{equation}
{\cal L} = \partial_\mu\phi\,\partial^\mu\phi + \mu \,m^2 \phi^2 - \partial_\mu\psi\,\partial^\mu\psi- f(\phi,\psi)\,, 
\label{eq:defineghosts}
\end{equation}
where $\phi$ has the wrong sign kinetic term and is coupled to some non-ghost fields represented by $\psi$. (We employ the mostly positive metric signature.) Let us first discuss the situation where $\mu>0$. In the absence of the coupling to other fields (such as $\psi$), the ghost does not lead to any instability either classically or quantum mechanically, since up to an overall sign, the theory is equivalent to a regular massive free field. On the other hand, the coupling $f(\phi,\psi)$ allows a rapid energy transfer from the ghost to non-ghost sector, rendering the vacuum $<\phi>=0$ unstable. Quantum mechanically, the phase space available for decay is of infinite measure, leading to a divergent decay rate \cite{Cline:2003gs}. On the other hand, the instability can be mild in effective field theories (EFT) where some UV completion takes over at a cutoff scale. If the mass of the ghost is above the cutoff scale, it will not be excited within the regime of validity of the EFT, thus is not a physical degree of freedom but an artifact of the low energy truncation (See Ref.~\cite{Carroll:2003st} for an example in the context of cosmology). 

Let us now turn to the opposite regime $\mu<0$, where the $\phi$ field is a tachyonic ghost and even the non-interacting theory is classically unstable due to the exponential growth with the imaginary mass. In this case, the EFT picture is also prone to the instability, as making the mass parameter $m$ large corresponds to a faster decay (up to the EFT cutoff). For these reasons, a tachyonic ghost appears to be harder to ``exorcise'' and is  typically not considered in the literature. 

A tachyonic instability occurs when the mass term of the field has the wrong sign and can actually correspond to a physical phenomenon. For instance, self-gravitating configurations have been long known to be unstable. The tendency of dust to clump together was observed more than a century ago by Jeans in Ref.~\cite{Jeans1902}, in the context of Newtonian gravity. This trend is also a feature present in General Theory of Relativity (GR), and most notably, its application to cosmology provides the current understanding of large scale structure formation in the Universe \cite{Peebles:1982ff,Blumenthal:1984bp,Davis:1992ui}. In the field theory framework, Jeans instability is a tachyonic instability, which can be traced back to a negative squared-mass contribution to the matter dispersion relation that dominates over the usual momentum piece in the region $k<k_J$, where $k_J$ is the characteristic wavenumber for the mode with a vanishing dispersion relation. In Minkowski background, it leads to an exponential growth of perturbations, while in an expanding background, the Hubble friction slows it down to a power-law evolution (see e.g. \cite{Mukhanov:2005sc}). This classical instability is conceptually well understood, it is physical and can be kept under control.

The goal of the present paper is to argue that a tachyonic ghost that emerges only far in the IR should not be cause of concern for quantum stability. It admits a classically equivalent ghost-free representation and the only persistent feature is a classical tachyonic instability. To illustrate this point in the simplest possible setting, we will focus on a very conservative matter sector consisting of a canonical scalar field minimally coupled to GR. We will first show in Sec.~\ref{sec:scalarexample} that the scalar field perturbations can become ghost-like in the IR around simple (non-trivial) backgrounds. We will then introduce in Sec.~\ref{sec:fluid} fluid description and in Sec.~\ref{sec:example-canonical} a canonical transformation that leads to a ghost-free reformulation in which velocity perturbations exhibit a classical tachyonic instability. Finally, we will demonstrate that this instability is equivalent to a Jeans instability. Sec.~\ref{sec:discussion} contains a discussion of our results. In particular we argue there that certain types of modified gravity theories are expected to exhibit apparently ghost-like instabilities of the type we identify here.

\section{IR ghost from matter coupled to GR}
\label{sec:scalarexample}

In order to illustrate our previous arguments in a concrete theory, we consider the Einstein-Hilbert action, along with a massless canonical scalar field coupled minimally to gravity,
\begin{equation}
 S=\int d^4x \sqrt{-g} \left[\frac{M_p^2}{2}\left(R-2\,\Lambda\right)
-\frac{1}{2}\,\partial_\mu\phi\,\partial_\nu\phi\,g^{\mu\nu}\right]\,.
\label{eq:example-action}
\end{equation}
Within the context of this theory, the effect that we are after occurs only in special background configurations where the gradient of the scalar field is space-like. This can be achieved by the background value
\begin{equation}
\phi =  M_p\,\sigma \,x\,,
\end{equation}
where $\sigma$ has a mass dimension 1 and as it corresponds to a constant gradient along the $\hat{x}$ direction, it acts as a source of shear. Thus, the background metric needs to be anisotropic. The simplest such metric is the Bianchi type I background with residual axisymmetry, 
\begin{equation}
ds^2= -dt^2 +a(t)^2dx^2+b(t)^2 (dy^2+dz^2)\,.
\end{equation}

By varying the action with respect to $g_{\mu\nu}$ and $\phi$, the background equations of motion can be calculated as:
\begin{eqnarray}
{\bf (i)}~~3\,H^2 &=& 3\,h^2+\frac{\sigma^2}{2\,a^2}+\Lambda\,,\nonumber\\
{\bf (ii)}~~2\,\dot{H} &=& -6\,h^2-\frac{\sigma^2}{3\,a^2}\,,\nonumber\\
{\bf (iii)}~~\dot{h}&=&-3\,h\,H+\frac{\sigma^2}{3\,a^2}\,,
\label{bgeq-GR}
\end{eqnarray}
where we defined the average expansion rate and the shear scalar as
\begin{equation}
H \equiv \frac{1}{3}\left(\frac{\dot{a}}{a}+2\,\frac{\dot{b}}{b}\right)\,,
\qquad
h \equiv \frac{1}{3}\left(\frac{\dot{a}}{a}-\frac{\dot{b}}{b}\right)\,,
\end{equation}
respectively.
We remark that the Eq.(\ref{bgeq-GR}) are connected through the contracted Bianchi identities:
\begin{equation}
 \partial_t {\bf (i)} 
  -3\,H\,{\bf (ii)}+6\,h\,{\bf (iii)}=0\,.
\end{equation}

Our goal is to verify the perturbative stability of this background. We thus introduce perturbations around the axisymmetric background, by
\begin{eqnarray}
&&g_{00}=-1-2\Phi\,,\qquad\qquad
g_{0x}=a\,\partial_x\chi\,,\qquad\qquad\qquad\qquad\;\;
g_{0I} = b(\partial_IB+\epsilon_I^J\partial_J B_{\rm odd})\,,\nonumber\\
&&g_{xx}=a^2(1+\psi)\,,\qquad
g_{xI}=a\,b\,\partial_x(\partial_I \beta+\epsilon_I^J\partial_J \beta_{\rm odd})\,,\qquad
g_{IJ} = b^2\left[\delta_{IJ}(1+\tau)+\partial_I\partial_J E+\partial_{(I}\epsilon_{J)}^K\partial_K E_{\rm odd}\right]\,,
\label{eq:perturbations}
\end{eqnarray}
while the scalar field is decomposed as
\begin{equation}
\phi = M_p\,\sigma(x+\partial_x\varphi)\,.
\end{equation}
In the following, we exploit the diffeomorphism invariance to fix the gauge as $\tau=\beta=E=E_{\rm odd}=0$.

The perturbations fall into two categories: odd modes which transform as a 2D vectors under rotations around the $\hat{x}$ axis, and even modes which transform as 2D scalars. After fixing the gauge, the odd sector consists of two degrees of freedom, $B_{\rm odd}$ and $\beta_{\rm odd}$. Out of these, $B_{\rm odd}$ does not have any time derivatives and its equation of motion can be immediately solved. Reducing the action by using the solution, then using the rescaled field $\beta_{\rm odd}/(b\,p)$ for canonical normalization, the dispersion relation for the odd mode can be found as
\begin{equation}
\omega^2_{\rm odd} = p^2 -\frac{9\,p_T^2(2\,p^2-3\,p_T^2)\,h^2}{p^4}+\frac{p_T^2\,\sigma^2}{p^2a^2}\,,
\end{equation}
where the physical momenta in the longitudinal and transverse directions are defined as
\begin{equation}
p_x \equiv \frac{k_x}{a}\,,\qquad
p_T \equiv \frac{k_T}{b}=\frac{\sqrt{k_y^2+k_z^2}}{b}\,,
\end{equation}
while the amplitude of the physical momentum is simply
\begin{equation}
p^2 \equiv p_x^2+p_T^2\,.
\end{equation}
Although there can be regimes where the frequency is complex, the kinetic coefficient of this mode is manifestly positive.

Let us turn now to the even sector. After fixing the gauge, there are 5 even degrees of freedom. Out of these $\Phi$, $\chi$ and $B$ are non-dynamical, which can be integrated out.  
The resulting reduced action now contains two dynamical degrees of freedom $\psi$ and $\varphi$ although they are coupled to each other. For this discussion, we only study the kinetic coupling. The relevant terms of the quadratic action in Fourier space are:
\begin{equation}
S = \frac{M_p^2}{2}\,\int d^3k\,dt\,a\,b^2\left[K_{11}\,\vert\dot{\psi}\vert^2 + K_{22}\,\vert\dot\varphi\vert^2+K_{12}\,(\dot\varphi^\star\,\dot\psi+\dot\psi^\star\,\dot\varphi)+\cdots\right]\,,
\end{equation}
where $K_{mn}$ can be seen as the components of a $2\times2$ matrix. To obtain the information about the signs of the coefficients of the kinetic terms in the diagonal basis, it is sufficient to look at the following combinations:
\begin{align}
\kappa_1^{\rm even} &= \frac{\det K}{K_{11}}= \frac{p_T^4p_x^2 \sigma^2a^2}{p_T^4 +\frac{2\,p^2\sigma^2}{a^2}}\,,\nonumber\\
\kappa_2^{\rm even} &= K_{11} = \frac{(h-H)^2 [p_T^4+2\,p^2 (\sigma^2/a^2)]}{2\,\left[(p_T^2-2\,p_x^2)h+2\,p^2H\right]^2 + 4\,(\sigma^2/a^2)\left[(p_T^2-3\,p_x^2)h^2+4\,p_T^2h\,H+(4\,p_T^2+3\,p_x^2)H^2\right]}\,.
\end{align}
Although the first eigenvalue is manifestly positive, the second one can be negative depending on the evolution and the value of the momentum for the mode. The negative contribution becomes important when the mode is aligned with the gradient of the scalar, i.e. $p_T \ll |p_x|\simeq p$. In this regime, the kinetic term is:
\begin{equation}
\left.\kappa_2^{\rm even}\right\vert_{p\simeq p_x} = \frac{(h-H)(\sigma^2/a^2)}{4\,p_x^2(h-H)-6\,(h+H)(\sigma^2/a^2)}\,,
\end{equation}
which indicates that for $p \simeq p_x$, the mode is a ghost if 
\begin{equation}
k_x^2 < \frac{3(h+H)\sigma^2}{2\,(h-H)}\,.
\label{eq:criticalmomentum}
\end{equation}
Thus, if the evolution allows $(h+H)/(h-H)>0$ to hold, there are always ghost modes with momenta $p_T^2 \ll p_x^2 < \sigma^2/a^2$. Noting that the kinetic coefficient $\kappa_2^{\rm even}$ has the same sign as the quantity $H^2-h^2$ at low momenta, we see from the first of Eq.~\eqref{bgeq-GR} that the only option to open up this regime of evolution is to have a sufficiently negative cosmological constant that can dominate over the energy density of the matter field. In this branch of  evolution the shear $h$ is positive whereas the average expansion rate $H$ eventually becomes negative, with $h > -H > 0$. In terms of the evolution of individual coordinates, the $x$ direction undergoes an accelerated expansion while the $y$ and $z$ directions contract. Thus, as the matter contribution in the Friedmann equation redshifts away, the difference $h^2-H^2$ converges to a constant determined by $\Lambda$, leading to an ever decreasing ratio $(h-H)/(h+H)$. The latter behavior indicates that as evolution goes on, fewer modes satisfy the inequality \eqref{eq:criticalmomentum}. Similarly, as the $y$ and $z$ directions contract, the physical momentum corresponding to these directions increase and the $x$ component of the physical momentum dominates for fewer modes.

Thus, for a very conventional example of a canonical scalar field minimally coupled to General Relativity, we have shown that the scalar field perturbations can be ghost-like. 

\section{Fluid analogue}
\label{sec:fluid}
In the previous section, we have shown that the matter perturbation can appear to be a ghost in the infrared. As we will show shortly, this is an artifact of choosing the field perturbation as the variable. Before choosing a variable with physical interpretation, we will change our perspective to a fluid description in this Section. 

The stress-energy tensor of a massless, canonical scalar field is
\begin{equation}
T_{\mu\nu} = \partial_\mu\phi \partial_\nu\phi- \frac{1}{2}\left(g^{\alpha\beta}\partial_\alpha\phi\partial_\beta\phi\right)g_{\mu\nu}\,.
\label{eq:tmunu-field}
\end{equation}
For an analogue fluid description, the stress-energy tensor is instead
\begin{equation}
T_{\mu\nu} = \rho \,u_\mu u_\nu+P(g_{\mu\nu}+u_\mu u_\nu)+\Pi_{\mu\nu}\,,
\label{eq:tmunu-fluid}
\end{equation}
where $\rho$, $P$ and $\Pi_{\mu\nu}$ are the energy density, pressure and shear tensor, respectively. The future-directed time-like vector $u_\mu$ satisfying $u_\mu u^\mu=-1$ is an eigenvector of the stress-energy tensor with the corresponding eigenvalue $-\rho$, and can be uniquely determined at non-linear level. The shear tensor is defined to be transverse and traceless i.e. 
\begin{equation}
\Pi_{\mu\nu}u^\nu = \Pi_{\mu\nu}g^{\mu\nu}=0\,.
\label{eq:pi-transverse}
\end{equation}
At the background level, we find the non-zero quantities as
\begin{equation}
\rho = \frac{M_p^2 \sigma^2}{2\,a^2}\,,\qquad
P = -\frac{\rho}{3}\,,\qquad
\Pi^x_{\;\;x} = \frac{4\,\rho}{3}\,,\qquad
\Pi^I_{\;\;J} = -\frac{2\,\rho}{3}\,\delta^I_J\,.
\end{equation}
Considering perturbations of the scalar field, one will inevitably generate perturbations in the energy-density, pressure and diagonal components of shear
\begin{equation}
\frac{\delta \rho}{\rho} = \frac{\delta P}{P} = \frac{\delta \Pi^x_{\;\;x}}{\Pi^x_{\;\;x}}=\frac{\delta\Pi^I_{\;\;I}}{\Pi^I_{\;\;I}}\Bigg\vert_{\rm no~sum}=2\,\partial_x^2\varphi-\psi\,,
\end{equation}
along with the spatial off-diagonal components
\begin{equation}
\delta\Pi^I_{\;\;x}=\frac{M_p^2\sigma^2}{b^2}\,\partial^I\partial_x\varphi+M_p^2\sigma^2\,\delta g^{xI}\,,\qquad
\delta\Pi^x_{\;\;I}=\frac{M_p^2\sigma^2}{a^2}\,\partial_I\partial_x\varphi\,.
\end{equation}
Defining the perturbed four-velocity as \footnote{We remark that a velocity component on the $y$--$z$ plane does not contribute to the stress-energy perturbations at linear order, hence it is taken to be zero without any effect on the result. This property can be traced back to the fact that the stress energy tensor in the $(t,y,z)$ directions, $T^\mathcal{A}_{\;\;\mathcal{B}}$, is proportional to $\delta^\mathcal{A}_\mathcal{B}$ up to linear order in $u^{y,z}$, where $\mathcal{A},\mathcal{B} = 0,2,3$.}
\begin{equation}
u^\mu = \left(1-\Phi, v, 0,0\right)\,,
\end{equation}
the transverse condition for the shear tensor \eqref{eq:pi-transverse} also generates perturbations of the following off-diagonal components:
\begin{equation}
\delta \Pi^x_{\;\;0} = -\Pi^x_{\;\;x} v\,,\qquad
\delta \Pi^0_{\;\;x} = \Pi^x_{\;\;x} a^2 \,\left(v+\frac{1}{a}\,\partial_x\chi\right)\,,\qquad
\delta \Pi^0_{\;\;I} = \Pi^J_{\;\;I}\,b\left(\partial_J B+ \epsilon^K_J\partial_J B_{\rm odd}\right)\,,
\end{equation}
where the velocity perturbation can be found to be
\begin{equation}
v = -\partial_x\partial_0 \varphi\,.
\label{eq:velocity}
\end{equation}

\section{Change of Variable}
\label{sec:example-canonical}

We now reconsider the model in Sec.~\ref{sec:scalarexample} and perform a canonical transformation.\footnote{One could perform a canonical transformation in the Hamiltonian formalism and then come back to the Lagrangian formalism by a Legendre transformation. In the present paper, we shall instead employ the technical procedure introduced in Appendix B of Ref.~\cite{DeFelice:2015moy}.  This makes it possible for us to perform the canonical transformation within the Lagrangian formalism. Needless to say, the two procedures are equivalent to each other.} Our goal is to change the nature of the ghost instability by defining a new variable. In order to identify a physical variable that is suitable for this task, we notice that the only perturbation in the fluid description that involves time derivative of the scalar field perturbation is the velocity perturbation \eqref{eq:velocity}. For later convenience, we choose the $x$ component of $\delta u_\mu$ to build our new variable:
\begin{equation}
\delta u_x = -a\,\partial_x (a\,\dot{\varphi}-\chi )\,.
\end{equation}
After using the decomposition \eqref{eq:perturbations}, the Lagrangian quadratic in even perturbations contains the following term
\begin{equation}
{\cal L}\ni \frac{M_p^2\sigma^2b^2}{2\,a} \left[\partial_x(a\,\dot{\varphi}-\chi )\right]^2\,,
\end{equation}
which is the only term that contains $\dot{\varphi}$ or its derivatives. Using Fourier series to expand the modes, we introduce the auxiliary variable $U$ by substituting 
\begin{equation}
\vert (a\,\dot{\varphi}-\chi )\vert^2 \quad
\Leftrightarrow \quad U\,(a\,\dot{\varphi}-\chi )^\star+ U^\star(a\,\dot{\varphi}-\chi )- |U|^2\,.
\end{equation}
The new field's equation of motion trivially gives $U= (a\,\dot{\varphi}-\chi )$ and it is related to the velocity perturbation through $\delta u_x=-a\,\partial_x U$. However, instead of using this solution, we can now integrate out the variable $\varphi$ which has become non-dynamical. This process is equivalent to the canonical transformation
\begin{equation}
U = \frac{2\,\Pi_\varphi}{M_p^2\sigma^2 a^2b^2p_x^2}\,,\qquad
\Pi_U = -\frac{1}{2}\,M_p^2\,\sigma^2a^2b^2p_x^2\,\varphi -\frac{2\,(H-h)}{a\,p^2}\,\Pi_\varphi \,,
\end{equation}
where the momentum conjugates of $\varphi^\star$ and $U^\star$ are
\begin{equation}
\Pi_\varphi = \frac{M_p^2\sigma^2a^2b^2p_x^2}{2}\,(a\,\dot{\varphi}-\chi)\,,
\qquad
\Pi_U= \frac{M_p^2\sigma^2b^2p_x^2}{4 p^2}\,\left(\Phi-\psi+2\,a\,\dot{U}\right)\,.
\end{equation}

We can then proceed to integrate out the non-dynamical fields $B$, $\chi$ and $\Phi$ and finally obtain an action with dynamical variables $(\psi,U)$. Doing a further field redefinition:
\begin{equation}
S_1 = \frac{p_T^2}{p^2}\,\psi\,,\qquad
S_2 = U-\frac{p_T^2}{2\,a\left[(2\,p_x^2-p_T^2)h-2\,p^2H\right]}\,\psi\,,
\end{equation}
the action reduces to
\begin{equation}
S = \frac{M_p^2}{2}\,\int d^3kdt\,a\,b^2\left[\kappa_1 \vert\dot{S}_1\vert^2 +\kappa_2 \vert\dot{S}_2\vert^2 - m_1^2 \vert S_1\vert^2 - m_2^2 \vert S_2\vert^2 -m_{12}(S_1 S_2^\star + S_1^\star S_2)\right]\,,
\end{equation}
where
\begin{equation}
\kappa_1 = \frac{p^4(H-h)^2}{2\,\left[2\,p^2H-(2\,p_x^2-p_T^2)h\right]^2}\,,\qquad
\kappa_2 = \frac{p_x^2\,\sigma^2}{p^2}\,.
\end{equation}
In other words, both kinetic terms are now manifestly positive. Although the two degrees of freedom are generically coupled in this field basis, they do decouple in the limit $p \simeq p_x \gg p_T$, where we have
\begin{align}
\kappa_1 &= \frac{1}{8}+ {\cal O}\left(\frac{p_T^2}{p_x^2}\right)\,,\qquad\qquad
\kappa_2 = \sigma^2+ {\cal O}\left(\frac{p_T^2}{p_x^2}\right)\,,\nonumber\\
m_1^2 &=\frac{p_x^2}{8}+ {\cal O}\left(\frac{p_T^2}{p_x^2}\right)\,,\qquad\qquad
m_{12} = {\cal O}\left(\frac{p_T^2}{p_x^2}\right)\,,\nonumber\\
m_2^2 & =\sigma^2 \left[
p_x^2 -2 (h-H)(5\,h+H) -\frac{2\,(2\,h+H)\,\sigma^2}{a^2(h-H)}-\frac{\sigma^4}{2\,a^4(h-H)^2}\right]
+ {\cal O}\left(\frac{p_T^2}{p_x^2}\right)\,.
\end{align}
In this limit, since the kinetic terms are time independent, the dispersion relations can be obtained simply by taking the ratio of the mass and kinetic terms:
\begin{equation}
\omega_1^2 \Big\vert_{p_x \gg p_T} = p_x^2\,,\qquad
\omega_2^2 \Big\vert_{p_x \gg p_T} = p_x^2 -\left[2 (h-H)(5\,h+H) +\frac{2\,(2\,h+H)\,\sigma^2}{a^2(h-H)}+\frac{\sigma^4}{2\,a^4(h-H)^2}\right]\,.
\end{equation}
We now see that in the regime of evolution discussed at the end of Sec.~\ref{sec:scalarexample} where we observed the ghost, i.e. $h > |H|$, the square-mass term becomes manifestly negative and the IR instability turns into a tachyonic instability. Just like Jeans instability, this growth affects only modes below a critical momentum.

 \section{Discussion}
\label{sec:discussion}

We have considered a simple canonical massless scalar field theory minimally coupled to General Relativity, in a special configuration where the scalar has a constant space-like gradient. The metric solution compatible with this setup is an axisymmetric Bianchi--I metric. We focused on a special branch of evolution which exists in the presence of a negative cosmological constant, where the universe overall rapidly contracts, while the direction coinciding with the gradient of the scalar expands. Studying the perturbations, we found that the kinetic term for the scalar perturbation becomes negative in the IR for modes with momenta parallel to the expanding direction. In short, the scalar field appears to be a ghost around this background.

Encountering a ghost  at high energies around sensible solutions is usually seen as an indication that the vacuum decays very rapidly. Finding a low energy ghost in our setting is by no means a sign of pathology for the theory. On the contrary, we demonstrated above that the ghost instability for the field perturbation actually has a physical interpretation. The scalar matter corresponds to an analogue fluid with constant energy density, pressure and shear. By performing a canonical transformation and changing the variable to the velocity perturbation, the ghost instability becomes classically equivalent at the level of the action to a Jeans instability for the new variable, effective below a characteristic wavenumber. This is another realization of the well-known instability of constant matter distributions against gravity. This demonstrates clearly our initial claim, that the appearance of a tachyonic ghost in the far infrared is not an indication for a catastrophic quantum instability.

In our example, the space-like gradient of the background scalar field plays an essential role in changing the sign of the original kinetic term. This allows usually suppressed couplings to survive at the level of the quadratic action and after all constraints are used, the negative contribution to the scalar field kinetic term can be seen to arise from the coupling between $\delta g_{00}$ and $\delta g_{0x}$ components, which is usually harmless for matter field with time-like gradient. In particular, the constraints now allow $\delta g_{00}$ to depend on the time derivative of the scalar field perturbation. 

Even though the instability here is limited to a finite duration of the evolution, and to an anisotropic background, the technical observation above indicates that similar situations can be encountered in more general setups where fields with space-like gradients are natural. First of all, the anisotropy of our background is solely due to the presence of a single scalar field, which will always pick a single direction. A natural construction to look for this effect would be modified gravity theories with broken diffeomorphism symmetry, where the gravitational analogue of the St\"uckelberg trick \cite{ArkaniHamed:2002sp} can be reversed by giving non-zero gradients to several scalar fields in the so-called unitary gauge. As a result, one ends up with multiple scalar fields each with non-zero gradients along different directions, thus retaining the isotropy of the space. In the context of these theories, our result indicates that a ghost instability constrained to the IR should be interpreted classically. Moreover, in theories where the constraint structure of GR is modified, a seemingly harmless, ``canonical'' matter field can have a dramatic impact. Although the behavior of matter perturbations in these types of theories is not completely understood due to the large number of degrees of freedom, we expect that there exist sensible backgrounds around which IR ghosts are present. Our study reveals that such IR ghosts can be as harmless as the classical Jeans instability and have a reasonable physical interpretation.

\acknowledgments
We thank Kazuya Koyama and Takahiro Tanaka for illuminating discussions at an early stage of this work. One of us (S.M.) thanks Antonio de Felice,  Rio Saitou, Misao Sasaki, Alexander Vikman and Yota Watanabe for useful discussions and for collaboration on a related subject. A.E.G. acknowledges support by STFC grant ST/L00044X/1. This work of S.M. was supported by Japan Society for the Promotion of Science (JSPS) Grants-in-Aid for Scientific Research (KAKENHI) No. 24540256, and by World Premier International Research Center Initiative (WPI), MEXT, Japan. The research of T.P.S. leading to these results has received funding from the European Research Council under the European Union's Seventh Framework Programme (FP7/2007-2013) / ERC grant agreement n. 306425 ``Challenging General Relativity''.  S.M. is grateful to colleagues at University of Nottingham, where this work was initiated and progressed during his visits, for warm hospitality.

\end{document}